\documentclass{PoS}

\usepackage{wrapfig,booktabs}

\newcommand{\order}{{\it O}}
\newcommand{\ba}{\begin{eqnarray}}
\newcommand{\ea}{\end{eqnarray}}
\newcommand{\be} {\begin{equation}}
\newcommand{\ee} {\end{equation}}

\title{Kaon semileptonic decays with $N_f=2+1+1$ HISQ fermions 
and physical light-quark masses}

\ShortTitle{$K\to\pi\ell\nu$ with $N_f=2+1+1$ HISQ fermions}

\author{
\speaker{E.~G\'amiz}$^{a}$,
A.~Bazavov$^b$,
C.~Bernard$^c$,
C.~DeTar$^d$,
D.~Du$^e$,
A.X.~El-Khadra$^f$,
E.D.~Freeland$^g$,
Steven~Gottlieb$^h$,
U.M.~Heller$^i$,
J.~Komijani$^j$, 
A.S.~Kronfeld$^{j,k}$,
J.~Laiho$^{e}$, 
P.B.~Mackenzie$^k$,
E.T.Neil$^l$,
T.~Primer$^m$, 
J.N.~Simone$^k$,
R.~Sugar$^n$,
D.~Toussaint$^m$,
R.S.~Van~de~Water$^k$,
and
Ran Zhou$^k$
 \\ \\
\llap{$^a$}
CAFPE and Departamento de F\'{\i}sica Te\'orica y del Cosmos,
Universidad de Granada, \hspace*{-0.4em}\thanks{Supported in part by the MINECO  
and Junta de Andaluc\'{\i}a (Spain).} Granada, Spain \\
\llap{$^b$}Department of Computational Mathematics, Science and Engineering
and Department of Physics and Astronomy, Michigan State University,
East Lansing, MI, USA\\
\llap{$^c$}Department of Physics, Washington University, St.~Louis, MO, USA \\
\llap{$^d$}Physics Department, University of Utah, Salt Lake City, UT, USA \\
\llap{$^e$}Department of Physics, Syracuse University, Syracuse, NY, USA\\
\llap{$^f$}Department of Physics, University of Illinois, Urbana, IL, USA \\
\llap{$^g$}Liberal Arts Department, School of the Art Institute of Chicago, Chicago, Illinois, USA\\
\llap{$^h$}Department of Physics, Indiana University, Bloomington, IN, USA \\
\llap{$^i$}American Physical Society, One Research Road, Ridge, NY, USA \\
\llap{$^j$}Institute for Advanced Study, Technische Univerit\"at M\"unchen, 
Garching, Germany\\
\llap{$^k$}Fermi National Accelerator Laboratory,\hspace*{-0.4em}
    \thanks{Operated by Fermi Research Alliance, LLC, under Contract
    No.~DE-AC02-07CH11359 with the US DOE.}~
Batavia, IL, USA \\
\llap{$^l$}Department of Physics, University of Colorado, Boulder, CO, USA\\
\llap{$^m$}Physics Department, University of Arizona, Tucson, AZ, USA\\
\llap{$^n$}Department of Physics, University of California, Santa Barbara,
USA\\

E-mail: \email{megamiz@ugr.es}}
\author{Fermilab Lattice and MILC Collaborations 
\hspace*{-0.4em}\thanks{Work supported in part by U.S. DOE under grants 
DE-FG02-91ER40628 (C.B.), 
DE-SC0010120 (S.G.),
DE-SC0010005 (E.T.N.), 
DE-FG02-13ER42001 \& DE-SC0015655 (A.X.K.),
DE-FG02-13ER41976 (T.P., D.T.), 
the U.S. NSF under grants
PHY10-034278 (C.D.),
PHY14-17805 (D.D., J.L.),
and PHY13-16748 (R.S.), the European Commission (E.G., A.S.K.) 
and the Intel Parallel Computing Center at Indiana University (R.Z). 
Computing resources were provided by the DOE, NSF and USQCD Collaboration.} 
}

\abstract{We discuss the reduction of errors in the calculation of the form factor 
$f_+^{K \pi}(0)$ with HISQ fermions on the $N_f=2+1+1$ MILC configurations from 
increased statistics on some key ensembles, new data on ensembles with lattice 
spacings down to $0.042~{\rm fm}$ and the study of finite-volume effects within 
staggered ChPT. We also study the implications for the unitarity of the CKM matrix 
in the first row and for current 
tensions with leptonic determinations of $\vert V_{us}\vert$. 
}

\FullConference{34th annual International Symposium on Lattice Field Theory\\
		24-30 July 2016\\
		University of Southampton, UK}

\begin{document}

\section{Introduction}

There exist several tensions involving the CKM matrix element 
$\vert V_{us}\vert$ as extracted from semileptonic kaon decays. First of all, 
there is a $\sim 2\sigma$ disagreement with unitarity in the first row of 
the CKM matrix, measured as the deviation from 0 of the quantity 
\ba
\Delta_u\equiv\vert V_{ud}\vert^2+\vert V_{us}\vert^2+\vert           
V_{ub}\vert^2 -1 = -0.00122(37)_{V_{us}}(41)_{V_{ud}}\,, \label{eq:unitarity}
\ea
where we have used the last published Fermilab Lattice/MILC result for the vector 
form factor at zero momentum transfer, 
$f_+(0)=0.9704(\pm0.33\%)$~\cite{Bazavov:2013maa,Gamiz:2013xxa}, the last 
experimental average of neutral and charged semileptonic $K$ decays 
in Ref.~\cite{Moulson:2014cra}, 
$\left.\vert V_{us}\vert f_+^{K\pi}(0)\right|_{exp} = 0.2165(\pm 0.18\%)$,   
$V_{ud}=0.97417(21)$ from superallowed nuclear $\beta$ decays~\cite{Hardy:2014qxa}, 
and we have neglected $|V_{ub}|$. The deviation does not change if one uses instead 
the values of the vector form factor from other recent lattice-QCD  
calculations~\cite{Carrasco:2016kpy,Boyle:2015hfa}.

\begin{center}
\begin{figure}[ŧh]
\begin{center}
\vspace*{-1.3cm}
\includegraphics[width=0.77\textwidth]{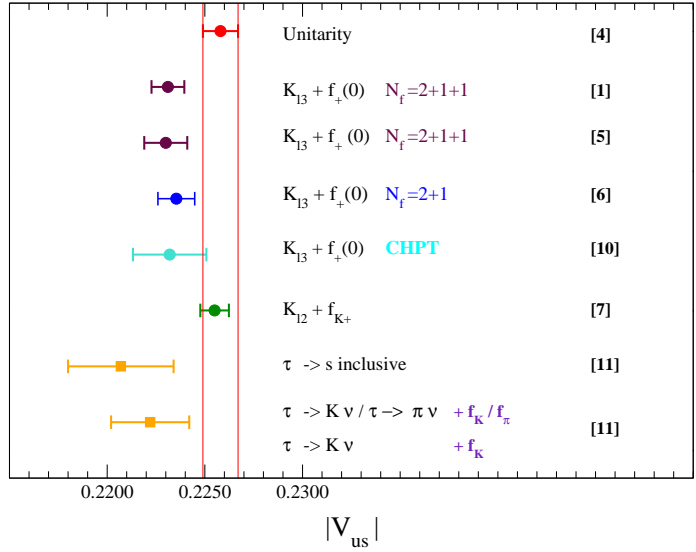}
\vspace*{-0.4cm}
\end{center}
\caption{Summary of recent $|V_{us}|$ determinations. \label{ResultsVus}}
\end{figure}
\end{center}

In addition, there is a $\sim2\sigma$ tension between the value of $\vert V_{us}\vert$ 
as obtained from leptonic and semileptonic decays~\cite{Rosner:2015wva}.  
The picture is complicated even further by including determinations of 
$\vert V_{us}\vert$  from hadronic $\tau$ decays, which traditionally yielded  
smaller values of $\vert V_{us}\vert$ than $K$ decays 
(and thus were in even more disagreement 
with unitarity) ---see Fig.~\ref{ResultsVus}. More recent results using 
lattice QCD for the hadronic vacuum polarization correlators 
and updated (preliminary) BaBar data, however, point 
towards values more compatible with unitarity~\cite{MaltmanLat2016}.

Checking the unitarity of the CKM matrix and the internal consistency of the 
Standard-Model description in the light sector is crucial for trying to unveil 
new-physics effects and put constraints on the scale of the allowed 
new physics~\cite{Gonzalez-Alonso:2016etj}.  
In order to perform even more stringent tests, it is necessary to reduce the error 
both on the experimental and lattice-QCD inputs entering determinations 
of $\vert V_{us}\vert$. For semileptonic decays, the lattice error 
on the vector form factor $f_+(0)$ is still the 
largest uncertainty. We summarize here our progress on reaching the objective 
of reducing the lattice-QCD error on $f_+^{K\pi}(0)$ to the 
$\sim 0.2\%$ level, which would match the current experimental uncertainty. 

\section{New simulation data}

The setup of our new work is the same as that used in our previous 
calculation~\cite{Bazavov:2013maa,Gamiz:2013xxa}. 
We obtain $f_+(0)$ from the relation $f_+(0)=f_0(0) = 
\frac{m_s-m_l}{m_K^2-m_\pi^2}\langle \pi(p_\pi)\vert S \vert K(p_K)\rangle$, 
where the vector form factor, $f_+$, and the scalar form factor, $f_0$, 
are defined by ($q=p_K-p_\pi$)
\ba
\langle \pi \vert V^\mu \vert K\rangle =  f_+(q^2) \left[p_K^\mu
+ p_\pi^\mu - \frac{m_K^2-m_\pi^2}{q^2}q^\mu\right]+ f_0(q^2)\frac{m_K^2-m_\pi^2}{q^2}q^\mu\,.
\ea 
We simulate directly at zero momentum transfer, $q^2\approx0$, by tuning the 
external momentum of the $\pi$ using partially twisted boundary conditions. 
We use the HISQ action for sea and valence quarks, simulating 
on the HISQ $N_f=2+1+1$ MILC configurations~\cite{HISQensembles}.

Our previous calculation~\cite{Bazavov:2013maa,Gamiz:2013xxa} used the HISQ 
action for both the sea and valence sectors, analyzed small 
lattice spacings down to $a\approx0.06~{\rm fm}$ and included ensembles with 
physical quark masses. The total error was dominated by statistics ---see 
Table~\ref{tab:errorbudget}.

\begin{table}[ht]
\begin{center}
\begin{tabular}{lcc}
\hline
 \hline
Source of uncertainty &  \begin{tabular}{c}Error $f_+(0)$ (\%)\\ Fermilab Lattice-MILC 2014 
[1]\\\end{tabular} &  \begin{tabular}{c}Error $f_+(0)$ (\%)\\ Preliminary 
2016\\\end{tabular} \\
\hline
Stat. + disc. + chiral inter. & $0.24$ & $\le 0.2$ \\
$m_s^{\rm val}\ne m_s^{\rm sea}$ & $0.03$ & $0.03$ \\
Scale $r_1$ & $0.08$  & $0.08$ \\
Finite volume & $0.2$  & $\sim 0$ \\
Isospin & $0.016$  & $0.016$ \\ 
\hline
Total Error & $0.33$ & $\le 0.22$\\
\hline\hline
\end{tabular}
\end{center}
\caption{\label{tab:errorbudget} Error budget for $f_+(0)$ in percent from our previous 
calculation in Ref.~\cite{Bazavov:2013maa} and error budget estimate from this work.} 
\end{table}

In order to reduce the statistical error, in the last two years we have generated new 
data on a number of key ensembles. Table~\ref{Ensembles} lists the final set of data 
as well as the data included in our 2014 analysis. We have more than doubled our 
statistics in one of the key ensembles for the chiral-continuum 
interpolation/extrapolation, the ensemble with $a\approx 0.09~{\rm fm}$ and physical 
quark masses (last line 
in Table~\ref{Ensembles} with $a\approx 0.09~{\rm fm}$), and the ensemble 
with smallest lattice spacing in our previous analysis, the one with 
$a\approx 0.06~{\rm fm}$ and $m_l=0.2m_s$ 
(first line in Table~\ref{Ensembles} with $a\approx 0.06~{\rm fm}$). 
In addition, in our new analysis we include the $a\approx 0.06~{\rm fm}$ 
ensemble with physical quark masses and a smaller lattice spacing, the 
ensemble with $a\approx 0.042~{\rm fm}$ and $m_l=0.2m_s$. These are the two last 
lines in Table~\ref{Ensembles}. Moreover, we have generated extra data 
in order to further investigate finite volume corrections as explained in the 
next section.

\begin{table}[ht]
\begin{center}\begin{tabular}{cccccccc}
\hline\hline
$\approx a({\rm fm})$ & $m_l/m_s$ & $m_\pi L$ & $N_{conf}\times N_{src}$ (2014) & 
$N_{conf}\times N_{src}$ (2016) & $am_s^{sea}$ & $am_s^{val}$ & \\
\hline
0.15   & 0.035 & 3.30 & $1000\times4$ & 
$1000\times4$ & 0.0647 & 0.0691 &  \\
\hline
0.12   & 0.2 & 4.54 & $1053\times8$ & $1053\times8$ & 0.0509  & 0.0535 & \\
       & 0.1 & 3.22 & - & $1020\times 8$ &  0.0507 & 0.053 & \\
       & 0.1 & 4.29 & $993\times 4$  & $993\times 4$ &  0.0507 & 0.053 & \\
       & 0.1 & 5.36 & $391\times 4$ & $1029\times 8$ &  0.0507 & 0.053 & \\
       & 0.035 & 3.88 & $945\times 8$ & $945\times 8$ &  0.0507  & 0.0531 & \\
\hline
0.09   & 0.2 & 4.50 & $773\times4$ & $773\times4$ &  0.037 & 0.038 & \\
       & 0.1 & 4.71 & $853\times4$ & $853\times4$ &  0.0363 & 0.038 &  \\
       & 0.035 & 3.66 & $621\times4$ & $950\times8$ &  0.0363 & 0.0363 & \\
\hline
0.06   & 0.2 & 4.51 & $362\times4$ & $1000\times8$ &  0.024 & 0.024 & \\
       & 0.035 & 3.95 & - & $692\times6$ &  0.022 & 0.022 & \\
\hline
0.042 & 0.2 & 3.23 & - & $432\times12$ &  0.0158 & 0.0158 & \\
\hline\hline
\end{tabular}
\end{center}
\caption{Parameters of the $N_f=2+1+1$ gauge-field ensembles used in
this work and details of the correlation functions generated.
$N_{\rm conf}$ is the number of configurations included in our analysis,
$N_{{\rm src}}$ the number of time sources used on each configuration,
and $L$ the spatial size of the lattice. The ensemble with 
$a\approx 0.12~{\rm fm}$, $m_l/m_s=0.1$ and $m_\pi L=3.22$ is not included 
in the main analysis, but used to analyze finite-volume effects. 
 \label{Ensembles}}
\end{table}

\begin{center}
\begin{figure}[ŧh]

\vspace*{-0.4cm}

\begin{minipage}[c]{.49\textwidth}
\begin{center}
2014

\vspace*{-0.2cm}
\includegraphics[width=1.\textwidth]{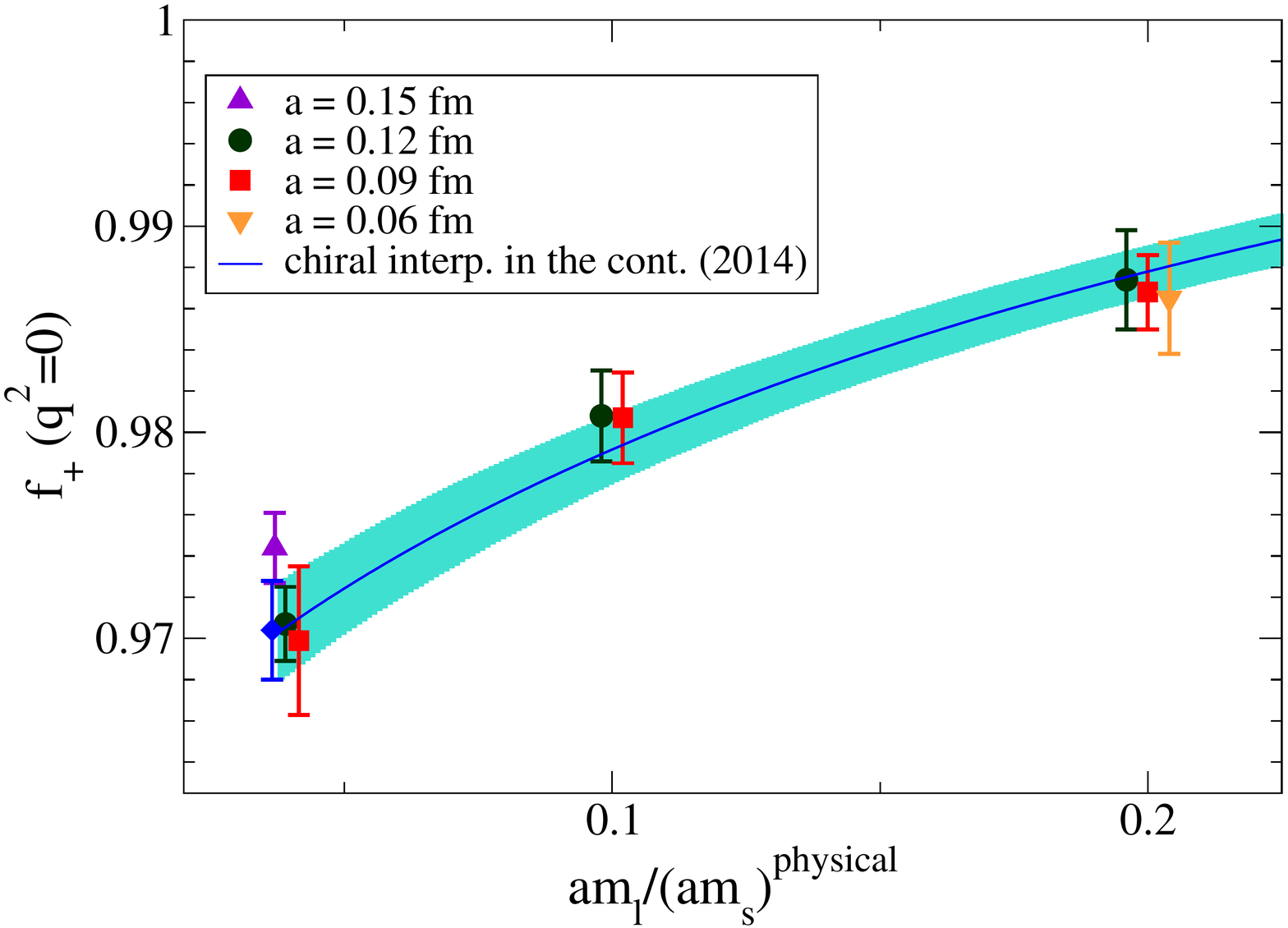}
\end{center}
\end{minipage}
\begin{minipage}[c]{.49\textwidth}
\begin{center}
2016

\includegraphics[width=1.\textwidth]{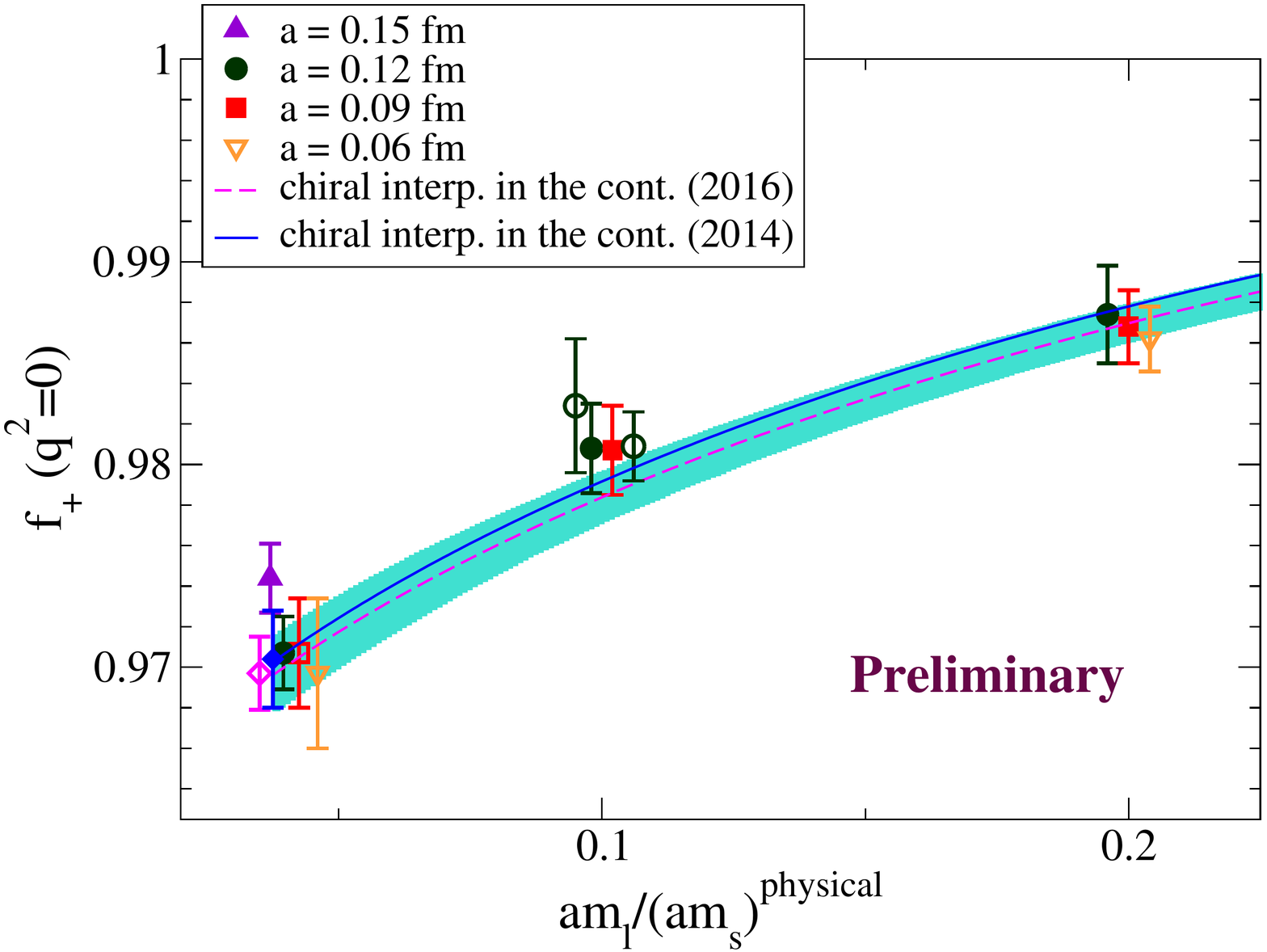}
\end{center}
\end{minipage}
\caption{Form factor $f_+(0)$ vs.~light-quark mass. 
Errors shown are statistical only, obtained from 500 bootstraps.
Different symbols and colors denote different lattice spacings, and open symbols 
correspond to either new ensembles or ensembles on which we have increased the 
statistics since our previous calculation in Ref.~\cite{Bazavov:2013maa}. 
The solid blue line is the interpolation in the light-quark mass, keeping 
$m_s$ equal to its physical value, and turning off all discretization effects, 
from Ref.~\cite{Bazavov:2013maa}. The dashed magenta line shows preliminary results 
for the same type of interpolation but including the new data generated since 2014.  
The blue and magenta diamonds are the corresponding interpolations at the 
physical point. 
The turquoise error band includes statistical, discretization and higher order 
chiral errors, as well as the uncertainty from some of the input parameters (see the 
text). Data at the same light-quark mass but different lattice spacing 
are off-set horizontally. 
\label{fig:extrapolation}}
\end{figure}
\end{center}

We block the data by four in all ensembles to avoid autocorrelations 
and follow the fit strategy already 
discussed in Refs.~\cite{Bazavov:2012cd} and~\cite{Lattice2012}. 
Preliminary results from those fits, except for the ensemble with 
$a\approx 0.042~{\rm fm}$ that has not yet been analyzed, are shown on the right hand 
side of Fig.~\ref{fig:extrapolation}. The (preliminary) errors in the data points 
are statistical only, obtained from 500 bootstraps. They range from $0.16\%$ to 
$0.38\%$.

In Fig~\ref{fig:extrapolation}, for comparison, we show both the results in 
our 2014 analysis (left plot) and the preliminary results we are reporting here 
(right plot). In the right plot we also show our previous chiral interpolation curve 
and the physical result (after performing the chiral-continuum
interpolation/extrapolation) together with the corresponding errors,  
in order to make the comparison easier. Solids symbols on that plot correspond to  
the ensembles 
for which our data set has not changed since 2014, while open symbols correspond 
to ensembles where we have increased the statistics or ensembles for which we did not 
have data previously. The dashed magenta line is the preliminary chiral interpolation 
including all the new data except for the ensemble with $a\approx 0.12~{\rm fm}$ and 
$m_\pi L=3.22$, and that with  $a\approx 0.042~{\rm fm}$. We use the same fit function 
as in our 2014 analysis, a NLO partially quenched staggered ChPT (PQSChPT) 
expression~\cite{Bernard:2013eya}, plus regular NNLO continuum ChPT 
terms~\cite{BT03}, plus extra analytic terms which parametrize higher order 
discretization and chiral effects:

\vspace*{-0.5cm}
\ba\label{eq:ChPTtwoloop}
f_+(0)  =   1   & + &   f_2^{{\rm PQSChPT}}(a) + K_1\,\sqrt{r_1^2a^2
\bar\Delta\left(\frac{a}{r_1}
\right) ^2} + K_3\,\left(\frac{a}{r_1}\right)^4 + f_4^{{\rm cont.}}\nonumber\\ 
 & + &   r_1^4(m_\pi^2-m_K^2)^2 \left[\,C_6+K_2\,\sqrt{r_1^2a^2\bar\Delta\left(\frac{a}{r_1}\right) ^2}
+ K_2'\,r_1²a^2\bar\Delta  + C_8\, m_\pi^2 + C_{10}\,m_\pi^4\right]\,,
\ea
where the constants $K_i$ and $C_i$ are fit parameters to be determined by the chiral 
fits using Bayesian techniques, $\bar\Delta$ is the average taste splitting
$\bar\Delta = \frac{1}{16}\left(\Delta_P+4\Delta_A+6\Delta_T+4\Delta_V+\Delta_I\right)$, and 
$r_1^2 a^2 \bar \Delta$ is a proxy for $\alpha_s^2 a^2$. More details about this fit 
function, priors used in the Bayesian approach and tests performed can be found in 
Refs.~\cite{Bazavov:2013maa,Gamiz:2013xxa}. 
The only change we have made to the fit function for this preliminary analysis is 
updating the values of the taste splitting, the relative scale $r_1/a$ and the pion 
decay constant, but the updates do not vary very much from the old values, and those 
changes have very little effect on the fit results. 

As explained in Refs.~\cite{Bazavov:2013maa,Gamiz:2013xxa}, the error in the physical 
values of $f_+(0)$ depicted in Fig.~\ref{fig:extrapolation} includes statistical 
(bootstrap), chiral extrapolation and discretization errors, as well as 
the uncertainty associated to those inputs treated as constrained fit parameters 
($\order(p^4)$ LEC's and  taste-violating hairpin parameters). 
The HISQ taste splittings are known precisely enough that their error has no impact on 
the final uncertainty. 

As Fig.~\ref{fig:extrapolation} shows, the addition of the new data only slightly 
changes the central value of $f_+(0)$, while the error is significantly reduced. 
These preliminary results seem to indicate that this error will be under 
$0.2\%$, which was our goal.

\section{Finite-Volume Effects}

The other dominant source of error in our 2014 calculation~\cite{Bazavov:2013maa}, 
as shown in Table~\ref{tab:errorbudget}, was the uncertainty due to finite-volume 
effects. For an estimate of the finite-volume error we compared  
data obtained on two ensembles with the same parameters but different 
spatial volumes, those with $a\approx 0.12$~fm and $m_l=0.1m_s$ and 
$m_\pi L = 4.29,5.36$ (third and fourth lines with $a\approx 0.12~{\rm fm}$ 
in Table~\ref{Ensembles}). The difference was about half of the statistical 
error, so we took the finite volume error to be the full size of the 
statistical error, $0.2\%$.

For our new analysis, we have increased the statistics on the largest $a\approx 0.12$~fm, 
$m_l=0.1m_s$ volume and generated data on a third, smaller, volume, so we have results 
for three different spatial volumes (with all other parameters fixed). The 
preliminary values we obtain for $f_+(0)$ on these three volumes are shown in 
Table~\ref{tab:FVchecks}. 
With the improved statistics the two largest volume results are basically 
the same, and the smaller volume gives a larger result but well within the statistical 
error. Disregarding the result on the $m_\pi L=3.22$ ensemble (which is smaller than all 
of the ensembles used in our main analysis), we conclude that finite-volume 
effects are smaller than our statistical errors, or 0.17\%. 

\begin{wraptable}{r}{8.5cm}
\begin{tabular}{cccc}
\hline\hline
$m_\pi L$ & 3.22 & 4.29 & 5.36 \\
\hline
$f_+(0)$ & 0.9829(33) & 0.9808(22)& 0.9809(17)\\
variation & +0.2\%  & $\sim0\%$ & $0\%$\\
\hline\hline
\end{tabular}
\caption{$f_+(0)$ on three different spatial volumes with $ a\approx 0.12~{\rm fm}$ and 
$m_l=0.1m_s$, and variation respect to the largest volume result. \label{tab:FVchecks}}
\end{wraptable}

To further reduce the finite-volume errors, we can study finite-volume effects 
systematically within the framework of ChPT, and 
then use the computed corrections to extrapolate our results to infinite volume. 
The use of partially twisted boundary conditions in the generation of 
the relevant correlation functions complicates the analysis in several ways. 
For example, at finite volume we need an extra form factors, $h_\mu$, to parametrize 
the weak current
\ba
\langle \pi^-(p')|V_\mu|K^0(p)\rangle = f_+(p_\mu+p'_\mu) + f_- q_\mu + h_\mu\,,
\ea
where the three form factors depend on the exact choice of twisting angles, not 
only on the value of $q^2$. At finite volume, then, the Ward-Takahashi identity 
that relates the matrix element of a vector current with that of a scalar current 
at zero momentum transfer leads to the following relation
\ba
f_+^V(0) = \frac{m_s-m_l}{\tilde m_K^2-\tilde m_\pi^2}\langle \pi(p_\pi)\vert 
S \vert K(p_K)\rangle -q_\mu h_\mu\,,
\ea
where the $\tilde m_P^2$ include one-loop finite-volume corrections~\cite{ChPTFV}. 

The authors of Ref.~\cite{ChPTFV} have performed the one-loop 
ChPT calculation of finite-volume effects for all the form factors 
involved in $K\to\pi\ell\nu$ decays, $f_+$, $f_-$ and $h_\mu$, including all choices 
of partially (and fully) twisted boundary conditions, both in the continuum and 
including the leading staggered effects. The main results are also 
reported at this conference~\cite{ChPTFVproc}. Both the partially quenched and 
full-QCD cases are studied in Ref.~\cite{ChPTFV}.  
When applying the one-loop formulae in Ref.~\cite{ChPTFV} to all ensembles included in
our calculation,~\footnote{We exclude here the ensemble with $0.12~{\rm fm}$, 
$m_l=0.1m_s$ and $m_\pi L=3.22$ because it is not used in the main 
analysis.} we conclude that the dominant finite-volume effects in our data
are under 0.05\% and higher order effects can be safely neglected. The one-loop
corrections are added before the chiral+continuum extrapolation and any remaining
finite-volume corrections are negligible.  

\section{Conclusions}

We have described here how we address the main two sources of uncertainty in our 
previous calculation of the vector form factor $f_+(0)$: 
statistics+discretization+chiral 
interpolation and finite volume effects. We summarize in the third column of 
Table~\ref{tab:errorbudget} our preliminary error budget for this quantity. We expect 
the final error to be reduced from $0.33\%$ to $\sim 0.2\%$, matching the level of 
precision of the average of experimental measurements, which includes the uncertainty 
from strong and electromagnetic isospin corrections.  

With this reduction in the error of $f_+(0)$, using the same experimental input and 
assuming that the central value does not vary, the unitarity test in 
Eq.~(\ref{eq:unitarity}) would become
\ba
\Delta_u\equiv\vert V_{ud}\vert^2+\vert V_{us}\vert^2+\vert           
V_{ub}\vert^2 -1 = -0.00122(27)_{V_{us}}(41)_{V_{ud}}\,, 
\ea 
which is $\sim 2.5\sigma$ away from the unitarity prediction with an error 
dominated by the uncertainty on $\vert V_{ud}\vert$. This will make revisiting the 
determination of $|V_{ud}|$ a priority for CKM tests. Again assuming the central 
value of the form factor does not change in our final analysis, 
the reduction of the error in $f_+(0)$ would increase the tension between the leptonic 
and semileptonic determinations of $\vert V_{us}\vert$ to the $2.7 \sigma$ level. 
Further, the semileptonic determination of $\vert V_{us}\vert$ would become more 
precise than the leptonic determination from $f_K$, and 
as precise as the leptonic determination using the ratio 
$f_K/f_\pi$ but without the need of the $\vert V_{ud}\vert$ input.

\end{document}